\def\A{\leavevmode\setbox0\hbox{A}\lower1.4ex\hbox to\wd0
{\hss`}\kern-.9\wd0A}
\def\E{\leavevmode\setbox0\hbox{E}\lower1.4ex\hbox to\wd0
{\hss`\/}\kern-.9\wd0E}
\def\a{\leavevmode\setbox0\hbox{a}\lower1.4ex\hbox to\wd0
{\hss`\/}\kern-\wd0a}
\def\e{\leavevmode\setbox0\hbox{e}\lower1.4ex\hbox to\wd0
{\hss`\/}\kern-\wd0e}
\newcommand{\be}{\begin{equation}}
\newcommand{\ee}{\end{equation}}
\newcommand{\ba}{\begin{array}}
\newcommand{\ea}{\end{array}}
\newcommand{\beqn}{\begin{eqnarray}}
\newcommand{\eeqn}{\end{eqnarray}}

\font\symb=msam7
\def\znakr{\raise1.5pt\hbox{\symb\char66\kern-2pt\char74}}
\def\znakl{\raise1.5pt\hbox{\symb\char73\kern-2pt\char67}}

\def\normalsize{
\setlength{\textheight}{23cm}
\setlength{\textwidth}{15cm}
\setlength{\topmargin}{-2.0cm}
\setlength{\hoffset}{-0.5cm}
\setlength{\leftmargin}{-1cm}
\setlength{\rightmargin}{2.0cm}}

\documentstyle[12pt]{article}
\normalsize

\begin{document}
\title{On Notivarg Propagator}
\author{ M. Bakalarska, W. Tybor\thanks{Supported by
\L{}\'od\'z University Grant No. 505/581} \\
Department of Theoretical Physics \\
University of \L \'od\'z \\
ul.Pomorska 149/153, 90-236 \L \'od\'z, Poland}
\date{}
\maketitle
\setcounter{section}{0}
\setcounter{page}{1}
\begin{abstract}
\noindent
The covariant propagator of the notivarg is found.
It has \\
the Feynmann - like form.
\end{abstract}

\newpage
{\bf 1.} The essential difficulty, we are faced with on calculating
the gauge field propagator, is that the Lagrangian operator of the
gauge theory is singular.The origin of the singularity is the gauge
freedom of the theory. To remove it we impose the suitable gauge
conditions. That allow us to obtain an effective action leading to
the regular Lagrangian operator. The propagator is defined as the
inverse operator to the Lagrangian one. The method works very well,
for example, in the electrodynamics. It has been verified in the
theory of the notoph [1,2]. However, the method is rather troublesome
if the gauge field is a many component tensor. Such a situation is in
the case of the notivarg that is described by the field with symmetry
properties of a Riemann tensor.\\
We can follow another way to obtain the form of propagators.
We consider the interaction of the gauge field with an external
current that obeys some conservation law ensuring the gauge invariance
for the interacting term. In the fixed Lorentz frame we perform the 
canonical analysis of the theory. Solving constraints we obtain the
physical Hamiltonian. Its free part describes only the physical 
degrees of freedom. The interaction part is a sum of two terms. 
The first one describes the interaction of the physical components 
of the gauge field with the current components that are not restricted
by the current conservation law. The second one describes the instant 
interactions, bilinear in the currents (for example, the Coulomb term 
$\rho\frac{1}{\Delta}\rho$ in the electrodynamics). Using the standard 
methods of the S - matrix formalism, with the help of this Hamiltonian,
we can calculate the amplitude of the current - current interaction,
squard in the coupling constant. On the other hand, the amplitude 
can be calculated using the covariant propagator. Taking into account 
the Lorentz properties of the field and the current and the law of the 
current conservation, we can predict the general form of the covariant 
propagator. The result of calculations must be independent of the 
calculation method. So, we can verify correctness of our prediction 
of the form of the covariant propagator.

{\bf 2.} The notivarg field $K^{\mu\nu\alpha\beta}$ interacting with
the external Weyl current $j^{\mu\nu\alpha\beta}$ is described by
the Lagrangian density [3]
\be
  {\cal L} = - (\partial_\sigma K^{\sigma\nu\alpha\beta})^{2} + 
  (\partial_\sigma {K^{\sigma\nu\alpha}}_\nu)^{2} + 
  \frac{1}{4} j^{\mu\nu\alpha\beta} K_{\mu\nu\alpha\beta},
  \label{1}
\ee
where $K^{\mu\nu\alpha\beta}$ has the symmetry properties of a 
Riemann tensor $K^{\mu\nu\alpha\beta} = K^{\alpha\beta\mu\nu} = - 
K^{\nu\mu\alpha\beta}$, $\varepsilon_{\mu\nu\alpha\beta} 
K^{\mu\nu\alpha\beta} = 0$. The Weyl current $j^{\mu\nu\alpha\beta}$
is the Riemann tensor obeying
\be
  {j^{\mu\nu\alpha}}_\nu = 0.
\ee
The conservation law for the Weyl current is 
\be
  \partial_\mu \partial_\alpha j^{\mu\nu\alpha\beta} = 0.
  \label{3}
\ee
The action determined by the Lagrangian (\ref{1}) is invariant under 
the gauge transformations 
\begin{eqnarray}
  \delta K^{\mu\nu\alpha\beta} & = & \varepsilon^{\mu\nu\sigma\lambda}
  \varepsilon^{\alpha\beta\varphi\kappa} \partial_\sigma \partial_
  \varphi \omega_{\lambda\kappa} + g^{\mu\alpha} (\partial^\nu 
  \eta^\beta + \partial^\beta \eta^\nu) + \mbox{} \nonumber \\
  && \mbox{} + g^{\nu\beta} (\partial^\mu \eta^\alpha + \partial^\alpha
  \eta^\mu) - g^{\mu\beta} (\partial^\nu \eta^\alpha + \partial^\alpha
  \eta^\nu) - g^{\nu\alpha} (\partial^\mu \eta^\beta + \partial^\beta
  \eta^\mu) + \mbox{} \nonumber \\
  && \mbox{} - 2 (g^{\mu\alpha} g^{\nu\beta} - g^{\mu\beta}
  g^{\nu\alpha}) \partial_\sigma \eta^\sigma ,
\end{eqnarray}
where $\omega_{\alpha\beta} = \omega_{\beta\alpha}$ and $\eta_\alpha$ 
are gauge tensors. \\
In the covariant gauge 
\be
  {K^{\mu\nu\alpha}}_\nu = 0, \qquad  \partial_\mu \partial_\alpha 
  K^{\mu\nu\alpha\beta} = 0
\ee
the field equation has the form \footnote{In Ref. [3] the right hand
side of Eq.\ (\ref{7}) reads $- \frac{1}{2} j^{\mu\nu\alpha\beta}$.} 
\be
  \Box K^{\mu\nu\alpha\beta} = - \frac{1}{2} j^{\mu\nu\alpha\beta}.
  \label{6}
\ee 
{\bf 3.} Let us consider the exchange of the notivarg between two 
external currents. The general structure of the amplitude describing 
the process in the second order of the perturbation theory is 
\begin{eqnarray*}
  A & = & (-i)^{2} \left( \frac{a}{k^{2}} j^{\mu\nu\alpha\beta} (-k) 
  j_{\mu\nu\alpha\beta} (k) + \frac{b}{k^{4}} k_\mu j^{\mu\nu\alpha
  \beta} (-k) k^\sigma j_{\sigma\nu\alpha\beta} (k) + \mbox{}\right.\\
  &&\left. \mbox{} + \frac{c}{k^{6}} k_\mu k_\alpha j^{\mu\nu\alpha
  \beta} (-k)  k^\sigma k^\kappa j_{\sigma\nu\kappa\beta} (k) \right),
\end{eqnarray*}
where $a$, $b$, $c$ are number factors. The last term vanishes due to 
the conservation law (\ref{3}). The second one has, in fact, the 
structure of the first one due to the following identity for the
Weyl tensor 
\[
  j^{\mu\nu\alpha\beta} j_{\sigma\nu\alpha\beta} = \frac{1}{4} 
  \delta^\mu_\sigma j^{\kappa\nu\alpha\beta} j_{\kappa\nu\alpha\beta}.
\]
Assuming the following form of the notivarg propagator
\begin{eqnarray}
  D_{\mu\nu\alpha\beta, \sigma\lambda\gamma\delta} (k) & = & - 
  \frac{1}{k^{2}} \frac{1}{8} \left(g_{\mu\sigma} g_{\nu\lambda}
  g_{\alpha\gamma} g_{\beta\delta} + g_{\mu\lambda} g_{\nu\sigma}
  g_{\alpha\delta} g_{\beta\gamma} + \mbox{}\right. \nonumber \\
  && \mbox{} + g_{\mu\gamma} g_{\nu\delta} g_{\alpha\sigma}
  g_{\beta\lambda} +  g_{\mu\delta} g_{\nu\gamma} g_{\alpha\lambda}
  g_{\beta\sigma} - g_{\mu\lambda} g_{\nu\sigma} g_{\alpha \gamma}
  g_{\beta\delta} + \mbox{} \nonumber \\
  &&\left. \mbox{} - g_{\mu\sigma} g_{\nu\lambda} g_{\alpha\delta}
  g_{\beta\gamma} -  g_{\mu\gamma} g_{\nu\delta} g_{\alpha\lambda}
  g_{\beta\sigma} -  g_{\mu\delta} g_{\nu\gamma} g_{\alpha\sigma}
  g_{\beta\lambda}\right)
  \label{7}
\end{eqnarray}
we obtain the amplitude 
\be
  A = (-i)^{2} (- \frac{1}{8}) j^{\mu\nu\alpha\beta} (-k) 
  D_{\mu\nu\alpha\beta, \sigma\lambda\gamma\delta} (k) 
  j^{\sigma\lambda\gamma\delta} (k).  
\ee  
The number factor $-\frac{1}{8}$ follows from Eqs.\ (\ref{1}) and 
(\ref{6})
\[
  \frac{1}{4} j^{\mu\nu\alpha\beta} K_{\mu\nu\alpha\beta} 
  \rightarrow \frac{1}{4} j^{\mu\nu\alpha\beta} D_{\mu\nu\alpha\beta,
  \sigma\lambda\gamma\delta} (- \frac{1}{2} j^{\sigma\lambda\gamma
  \delta}).
\]
Using\\
 (i) the current conservation law (\ref{3}),\\
 (ii) the decomposition of the Weyl current  
\[
  j^{\mu\nu\alpha\beta} = (\tau^{ij}, \sigma^{ij}), \qquad
  i,j = 1, 2, 3;
\]
where $\tau^{ij}$ and $\sigma^{ij}$ are symmetric and traceless 
tensors defined by 
\begin{eqnarray*}
 && j^{0i0j} = \tau^{ij},\\
 && j^{0ijk} = \varepsilon^{jkp} \sigma^i_p,\\
 && j^{ijkl} = - (g^{ik} \tau^{jl} + g^{jl} \tau^{ik} - g^{il}
 \tau^{jk} - g^{jk} \tau^{il}).
\end{eqnarray*}
 (iii) the helicity decomposition of the symmetric traceless
 tensor $a^{ij}$
\[
  a^{ij} = a^{ij} (0) + a^{ij} (\pm 1) + a^{ij} (\pm 2)
\]
where
\begin{eqnarray*}
  a^{ij} (\pm 1) & = & - \frac{1}{\Delta} (\partial^i a^j_T + 
  \partial^j a^i_T),\\
  a^{ij} (0) & = & \frac{3}{2} (\frac{1}{\Delta} \partial^i \partial^j
  + \frac{1}{3} g^{ij}) a_L,\\
  a^i & \equiv & \partial_j a^{ji}, \\
  a_L & \equiv & \frac{1}{\Delta} \partial_i a^i,
\end{eqnarray*}
we obtain 
\begin{eqnarray}
  A & = & \frac{3}{2} \frac{\sigma_L (-k) \sigma_L (k)}{k^{2}} + 
  2 {\mid \vec{k} \mid}^{-4} \tau_{Ti} (-k) \tau^i_T (k) + \mbox{} 
  \nonumber \\
  &&\mbox{} + \frac{1}{4} k^{2} k^{-2}_{0}{\mid \vec{k} \mid}^{-2}
  \tau_{ij} ({\pm 2}, {-k}) \tau^{ij} ({\pm 2}, k).\label{9}
\end{eqnarray}
From the canonical analysis [4] we obtain the physical Hamiltonian 
density describing interaction 
\begin{eqnarray}
  {\cal H}_{int} & = & \sqrt{\frac{3}{2}} \varphi \sigma_L +
  \frac{1}{2} \sigma^n_m (\pm 2) \frac{1}{\Delta} \sigma^m_n (\pm 2)
  + \mbox{} \nonumber \\
  && \mbox{} + \frac{1}{8} (\frac{1}{\Delta} \partial^{0} \tau^{ij} 
  (\pm 2)) (\frac{1}{\Delta} \partial^{0} \tau_{ij} (\pm 2)) - 
  \frac{3}{8} \tau^{ij} (\pm 2) \frac{1}{\Delta} \tau_{ij} (\pm 2) + 
  \mbox{} \nonumber \\  
  &&\mbox{} - (\frac{1}{\Delta} \tau^i_T) (\frac{1}{\Delta} \tau_{Ti}).
  \label{10}
\end{eqnarray}
Two remarks are necessary: \\
1. The term (Eq.\ (34) in Ref. [4])
\[
  - \frac{1}{4} \sigma^n_m (\pm 2) \frac{1}{\Delta} \sigma^m_n (\pm 2)
\]
is wrong. The right form is 
\[
  - \frac{1}{2} \sigma^n_m (\pm 2) \frac{1}{\Delta} \sigma^m_n (\pm 2).
\]
2. The new field $\varphi = \sqrt{6} S_L$ is introduced because the free
physical Lagrangian has then the standard form 
\[
  {\cal L}_{free} = \frac{1}{2} (\partial_\mu \varphi)^{2}.
\]
We recall that $S_L$ is a scalar component of the symmetric traceless
tensor
\[
  S^{im} = - \frac{1}{4} ({\varepsilon^m}_{jk} K^{0ijk} + 
  {\varepsilon^i}_{jk} K^{0mjk}).
\]
The Hamiltonian density (\ref{10}) in the momentum space is 
\begin{eqnarray}
  {\cal H}_{int} & = & \sqrt{\frac{3}{2}} \sigma_L (-k) \varphi (k) + 
  \frac{1}{2} \left[ 2 {\mid \vec{k} \mid}^{-4} \tau_{Ti} (-k)
  \tau^i_T (k) + \mbox{}\right.  \nonumber \\
  &&\left. \mbox{} + \frac{1}{4} k^{2} k_{0}^{-2} {\mid \vec{k}
  \mid}^{-2} \tau_{ij} ({\pm 2}, {-k}) \tau^{ij} ({\pm 2},k)\right].
  \label{11}
\end{eqnarray} 
The current conservation law (\ref{3}) is included in
Eq.\ (\ref{11}).\\
With the help of this Hamiltonian, using standard methods of the 
S - matrix formalism [5], we can obtain the amplitude of the current - 
current interaction via one notivarg exchange. We get exactly the 
amplitude (\ref{9}). So, the Feynmann - like form (\ref{7}) of 
the notivarg propagator is confirmed. 

\bigskip   We are grateful to Prof. J. Rembieli\'nski for interesting 
discussion. 

\newpage

\end{document}